# Nanoscale Detection of Magnon Excitations with Variable Wavevectors Through a Quantum Spin Sensor

Eric Lee-Wong[1,2], Ruolan Xue[3,4], Feiyang Ye[1], Andreas Kreisel[5], Toeno van der Sar[6], Amir Yacoby[3,4], Chunhui Rita Du[1]

[1]Department of Physics, University of California San Diego, La Jolla, California 92093, USA.

[2]Department of NanoEngineering, University of California, San Diego, La Jolla, California 92093, USA.

[3]Department of Physics, Harvard University, 17 Oxford Street, Cambridge, Massachusetts 02138, USA.

[4]John A. Paulson School of Engineering and Applied Sciences, Harvard University, Cambridge, Massachusetts 02138, USA.

[5]Institute for Theoretical Physics, University of Leipzig, Brderstr.16, 04103 Leipzig, Germany.

[6]Kavli Institute of Nanoscience, Delft University of Technology, 2628CJ Delft, The Netherlands.

**Abstract:** We report the optical detection of magnons with a broad range of wavevectors in magnetic insulator $Y_3Fe_5O_{12}$ thin films by proximate nitrogen-vacancy (NV) single-spin sensors. Through multi-magnon scattering processes, the excited magnons generate fluctuating magnetic fields at the NV electron spin resonance frequencies, which accelerate the relaxation of NV spins. By measuring the variation of the emitted spin-dependent photoluminescence of the NV centers, magnons with variable wavevectors up to ~$5 \times 10^7$ m$^{-1}$ can be optically accessed, providing an alternative perspective to reveal the underlying spin behaviors in magnetic systems. Our results highlight the significant opportunities offered by NV single-spin quantum sensors in exploring nanoscale spin dynamics of emergent spintronic materials.

Control and manipulation of spin currents in an energy-efficient manner has been a central focus of modern spintronic research.[1] Magnons, bosonic type quasiparticles carrying quanta of spin angular momentum, are naturally relevant in this context due to their long coherence length, an extended lifetime, and reduced energy dissipation channels, offering remarkable opportunities in designing next-generation multifunctional spintronic devices.[2] To date, tremendous research efforts have been dedicated towards this end. Examples include magnon condensation,[3] spin-superfluidity,[4] interplay between spin waves and magnetic domain walls,[5–9] magnon-driven spin-torque oscillators,[10,11] and many others.[12] There are ongoing intense activities to investigate and understand the emergent magnonic systems as well as to create new ones. The success of these efforts relies simultaneously on advances in theory, material synthesis, and the development of new, sensitive measurement techniques.

The existing state-of-the-art efforts in probing magnons have been mainly focused on spin transport,[13] ferromagnetic resonance (FMR) spectroscopy,[14,15] and inelastic magnon-photon interactions, such as Brillouin light scattering measurements[16,17] and scanning x-ray transmission microscopy.[18] Due to the geometric restrictions and the optical diffraction limit, it is usually challenging to maintain nanoscale spatial resolution simultaneously with a broad detection range for magnon wavevectors.[16,19] This set of limitations inherently impose a hurdle to a comprehensive understanding of the underlying microscopic magnon-magnon interactions,[20,21] magnon thermalization,[22] and magnon Bose-Einstein condensation,[23,24] where exchange magnons play a dominant role in these situations.

Nitrogen-vacancy (NV) centers, optically-active atomic defects in diamond that act as single-spin quantum bits, are naturally relevant in this context due to their excellent quantum coherence, local spin addressability, and notable versatility in a wide temperature range.[25–27] Here, we employed NV singe-spin sensors[25] to perform nanoscale detection of magnons with a broad range of wavevectors generated in magnetic insulator $Y_3Fe_5O_{12}$ (YIG) thin films. The measured magnon spectrum can be well interpreted by the variation of the magnon band structure with film thicknesses and wavevectors. The sensitivity length scale and the measurable range of magnon wavevectors are mainly determined by the NV-to-sample distance, which can ultimately approach the tens-of-nanometer regime,[28,29] enabling a new opportunity to extract previously inaccessible information of nanoscale magnetic excitations in a variety of magnetic materials. Furthermore, the demonstrated coupling between NV centers and the exchange magnons also points to the possibility to develop NV-magnon-based hybrid quantum architectures for next-generation quantum information technologies.[30–32]

We start from discussing the NV measurement platform and device structure as illustrated in Fig. 1(a). A patterned diamond nanobeam[33] containing individually addressable NV centers is transferred on a YIG thin film grown on a $Gd_3Ga_5O_{12}$ substrate. The NV-to-sample distance typically lies in the range of ~100 nm in this study (see Supporting Information for details), ensuring nanoscale spatial sensitivity. A 600-nm-thick and 6-μm-wide Au stripline is fabricated on top of the YIG thin film to provide microwave control of the magnon excitations and the NV spin states. The negatively charged NV state has an $S = 1$ electron spin with a spin triplet ground state ($m_s = 0, \pm1$) as illustrated in Fig. 1(b). This three-level spin system can be optically read out by spin-dependent photoluminescence (PL), where the $m_s = \pm1$ spin states are more likely to be trapped by a non-radiative pathway (in the red wavelength range) through an intersystem crossing and back to the $m_s = 0$ ground state, yielding a significantly reduced PL intensity.[25] The exhibited spin-dependent PL sensitivity of NV centers provides a convenient way to probe the magnon excitations of a proximal YIG thin film, which will be discussed in detail later.

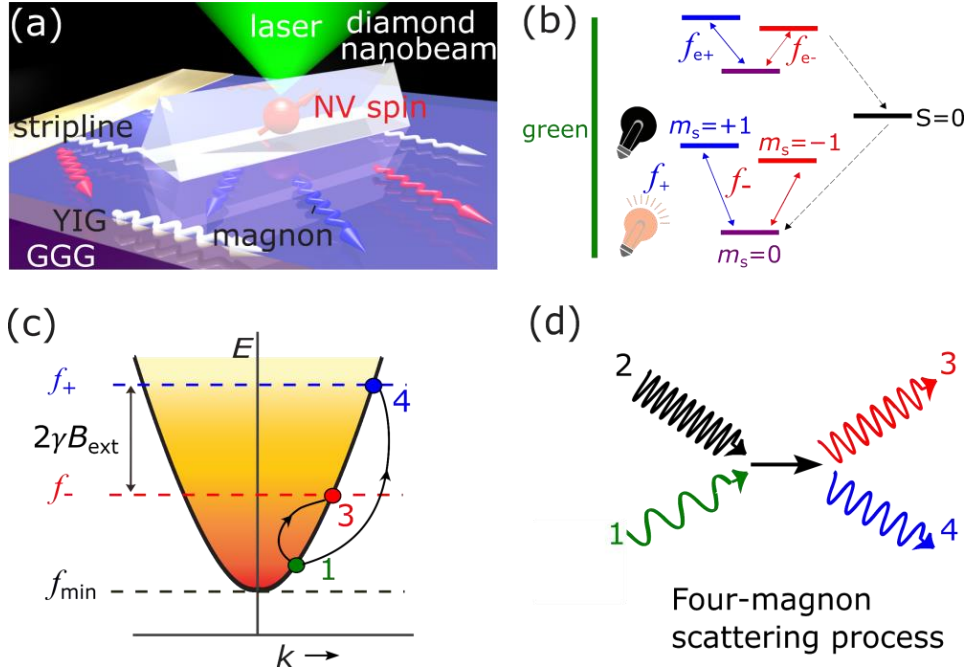

**Figure 1**: (a) Schematic of using an NV single-spin sensor contained in a diamond nanobeam to locally probe the magnons of a proximal YIG thin film. (b) Energy diagram of the $S = 1$ NV spin state, where the $m_s = \pm 1$ states fluorescence less strongly than the $m_s = 0$ state. (c) Sketch of the magnon dispersion and the magnon density. The generated magnons (green color dot) at a certain frequency lead to an enhanced magnon density at the NV ESR frequencies $f_\pm$ (red and blue dots) through the multi-magnon scattering processes. (d) Schematic of the four-magnon scattering process, where the scattering between magnon #1(green) and magnon #2 (black) generates magnon #3 (red) and magnon #4 (blue) with different frequencies and wavevectors.

Next, we briefly describe the coupling mechanism between magnons and an NV single-spin sensor. Figure 1(c) illustrates the sketch of the magnon dispersion and the magnon density of a YIG thin film which falls off as 1/Energy (1/$E$) as indicated by the fading color. With a moderate external magnetic field applied along the NV-axis, the NV electron spin resonance (ESR) frequencies $f_\pm$, corresponding to transitions between the $m_s = 0$ and the $m_s = \pm 1$ states, stay above the minimum of the YIG magnon energy $f_{\min}$. With microwave excitations, extra magnons with a certain wavevector and a frequency will be generated as denoted by the green dot. Due to the exchange interaction,[15] an excited magnon #1 will scatter with a thermal magnon #2, generating two new incoherent magnons (magnon #3 and magnon #4) with different wavevectors and frequencies as illustrated in Fig. 1(d). Energy and momentum are conversed during these processes. Continuously circulating the above four-magnon scattering processes will redistribute the magnon distribution and lead to the establishment of a new thermal equilibrium state with an enhanced magnon density at the frequencies $f_\pm$.[34–36] The increased magnetic fluctuations at the NV ESR frequencies $f_\pm$ will accelerate the NV relaxation from the $m_s = 0$ to the $m_s = \pm 1$ states, giving rise to a variation of the measured PL intensity.[35,36] In the following discussion, we assume that the change of the PL intensity is dominated by the variation of magnon density at a frequency $f_-$ in the low magnetic field regime ($f_- > f_{min}$). With a sufficiently large magnetic field ($f_- < f_{min} < f_+$), the change of the PL intensity will be mainly driven by the variation of magnon density at a frequency $f_+$.[35]

To generate magnons with a broad range of wavevectors, we first employed the nonlinear parametric excitation to generate exchange magnons in a 100-nm-thick YIG thin film. As illustrated in Fig. 2(a), parametric excitation harnesses the elliptically shaped precession of the magnetization. When a sufficiently large microwave field $B_{mw}$ with a frequency $f_{mw}$ is applied parallel to the out-of-plane component of the YIG magnetization, exchange magnons with high wavevectors ($\geq 1 \times 10^7$ m$^{-1}$) at a frequency $f_{mw}/2$ will be generated.[37–39] Figure 2(b) shows the magnon band structure of a 100-nm-thick YIG thin film (see Supporting Information for details), where the purple line marks the driving frequency $f_{mw}$ of the microwave field, the red line marks the NV ESR frequency $f_-$, the green line marks the frequency $f_{mw}/2$ of the parametrically excited magnons, and the blue line represents the band minimum $f_{min}$ of the 100-nm-thick YIG film. Note that $f_{min}$ is below the FMR frequency $f_{FMR}$ at wavevector $k = 0$ due to magnetostatic coupling.[40]

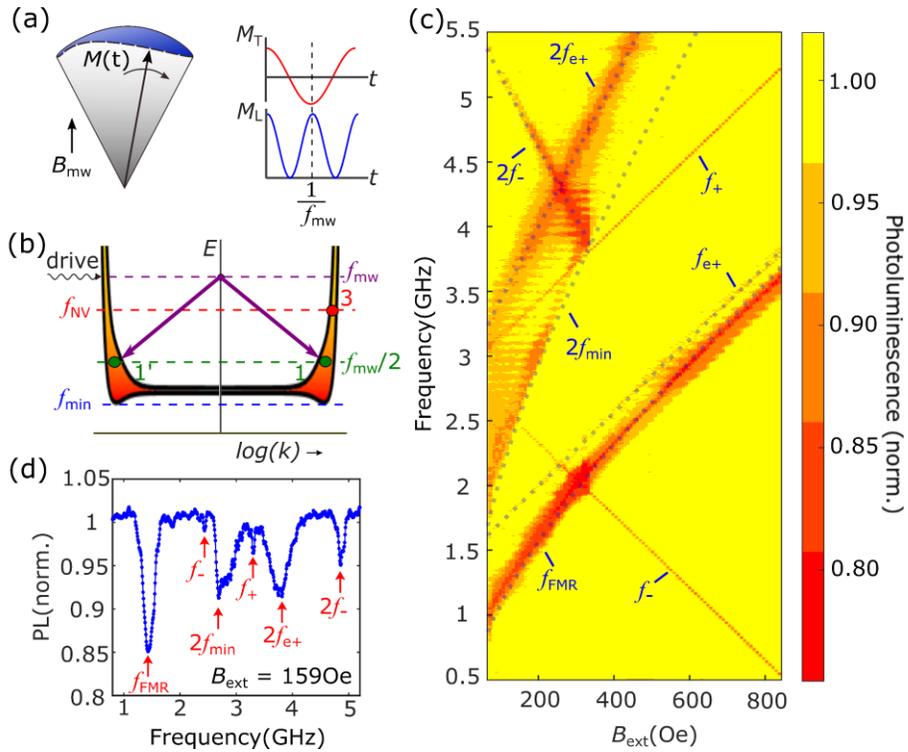

**Figure 2:** (a) Schematic of the parametric spin excitation: the magnetization precesses in an elliptical orbit under a parallel microwave pumping field with a frequency $f_{mw}$. The longitudinal magnetization $M_L$ oscillates at a frequency $f_{mw}$ and the transverse component $M_T$ oscillates at a frequency $f_{mw}/2$. (b) Sketch of the magnon band structure of a 100-nm-thick YIG thin film, where the location of the microwave drive frequency $f_{mw}$ (purple color), the half frequency of the drive field $f_{mw}/2$ (green color), the NV ESR frequency $f_{NV}$ (red color), and the minimal magnon band energy $f_{min}$ (blue color) are marked. The upper branch of the dispersion curve corresponds to the magnetostatic surface spin wave mode and the lower branch corresponds to the backward volume spin wave mode. The shaded area between the two curves represents the magnon density. (c) Normalized PL intensity of the NV center as a function of $B_{ext}$ and the microwave frequency. Semi-transparent dashed lines represent the calculated dispersion curves of individual magnon modes with the characteristic frequencies marked. They are in agreement with the observed features of the PL spectrum. (d) A linecut at $B_{ext} = 159$ Oe showing the corresponding dips of the measured NV PL intensity.

To perform optical detection of magnetic resonance (ODMR) measurements, a constant green laser excitation is applied to the NV center, and the emitted PL is monitored via a single-photon detector. An external magnetic field $B_{ext}$ is applied along the NV-axis, which makes a 61-degree angle relative to the normal of the film plane. The local microwave field $B_{mw}$ applied at the NV site is estimated to be 1.8 Oe characterized by the NV Rabi oscillation measurements (see Supporting Information for details). Figure 2(c) shows the normalized PL intensity as a function of microwave frequency $f_{mw}$ and the external magnetic field $B_{ext}$. The two straight lines from 2.87 GHz result from the expected decrease in NV fluorescence when $f_{mw}$ matches one of the NV ESR frequencies: $f_{\pm} = 2.87 \pm \gamma B_{ext}$, where $\gamma$ denotes the gyromagnetic ratio. A straight line with the same slope starting at 1.42 GHz comes from the NV ESR at the optically excited state: $f_{e+} = 1.42 + \gamma B_{ext}$. The NV fluorescence also decreases when $f_{mw}$ matches the calculated FMR frequency $f_{FMR}$ of the 100-nm-thick YIG film as marked by the curved dash line below $f_{e+}$. In addition to the above features that have been previously observed,[35,36,41,42] notably, groups of magnon excitations also emerge at higher frequencies with a threshold frequency following $2f_{min}$: twice of the minimal energy of the magnon band, exhibiting the hallmark of parametric excitation. For $B_{ext} = 159$ Oe, the measured $f_{min}$ equals 1.34 GHz, in agreement with the theoretical calculation (see Supporting Information for details). The measured PL intensity also significantly decreases when $f_{mw} = 2f_-$ and $f_{mw} = 2f_{e+}$. These can also be explained by the parametric spin excitation processes. As the frequency of the excited magnons lies exactly at the ground (excited) spin transition frequency $f_-$ ($f_{e+}$), magnons could directly couple with the NV spin, leading to accelerated NV relaxation rates with an improved optical addressability. This is unlike the situation of the four-magnon scattering processes discussed above, where the generated pair of magnons have different energies with the NV spin transition frequency. Enhancement of the magnon density at the ESR frequency is attributed to the circulation of the multi-magnon scattering processes, showing a reduced optical contrast. Figure 2(d) shows a linecut at $B_{ext} = 159$ Oe of the measured ODMR map, exhibiting clear dips at frequencies $f_{FMR}$, $2f_{min}$, $2f_-$, and $2f_{e+}$. Here, we note that previous work on probing the off-resonant NV-magnon coupling have been mainly focused on the uniform FMR mode or built on nanodiamond with multiple NV orientations and dramatically reduced spin coherence time .[31,36] Our results provide the first clear evidence to demonstrate the intrinsic coupling between exchange spin waves with an NV single-spin qubit.

To demonstrate the universality of our measurement technique, we varied the thickness and the dimensions of the YIG thin film to modify the magnon band structure and the associated wavevectors of the excited spin waves.[43] When the film thickness increases from 100 nm to 3 μm, the magnon band structure significantly changes due to the enhanced dipolar interaction as illustrated in Fig. 3(a). In this case, $f_{min}$ is significantly lower than $f_{FMR}$, and magnons can be confined along the film thickness direction, leading to a family of thickness modes with wavevectors: $k_n = n\pi/t$, where $t$ is the film thickness and the integer $n$ characterizes the mode number. Figure 3(b) shows the ODMR map of a 3-μm-thick YIG thin film, which exhibits richer magnon features in comparison to the 100-nm-thick film. In addition to the previously observed magnon features at a frequency $f_{FMR}$ and groups of parametric magnon excitations at frequencies $2f_{e\pm}$ and $2f_-$, we further observed the decrease of the PL intensity at $f_{mw} = 2f_{min}$ ($n = 1$), resulting from the parametric excitation of the $n = 1$ thickness mode. Moreover, at $f_{mw} = 2f_-/3$ and $f_{mw} = 2f_+/3$, magnon excitations also emerge, corresponding to the generation of spin waves at half integer multiples of the microwave drive frequency. Note that similar features have been observed in ferromagnetic NiFe films in the low magnetic field regime.[44]

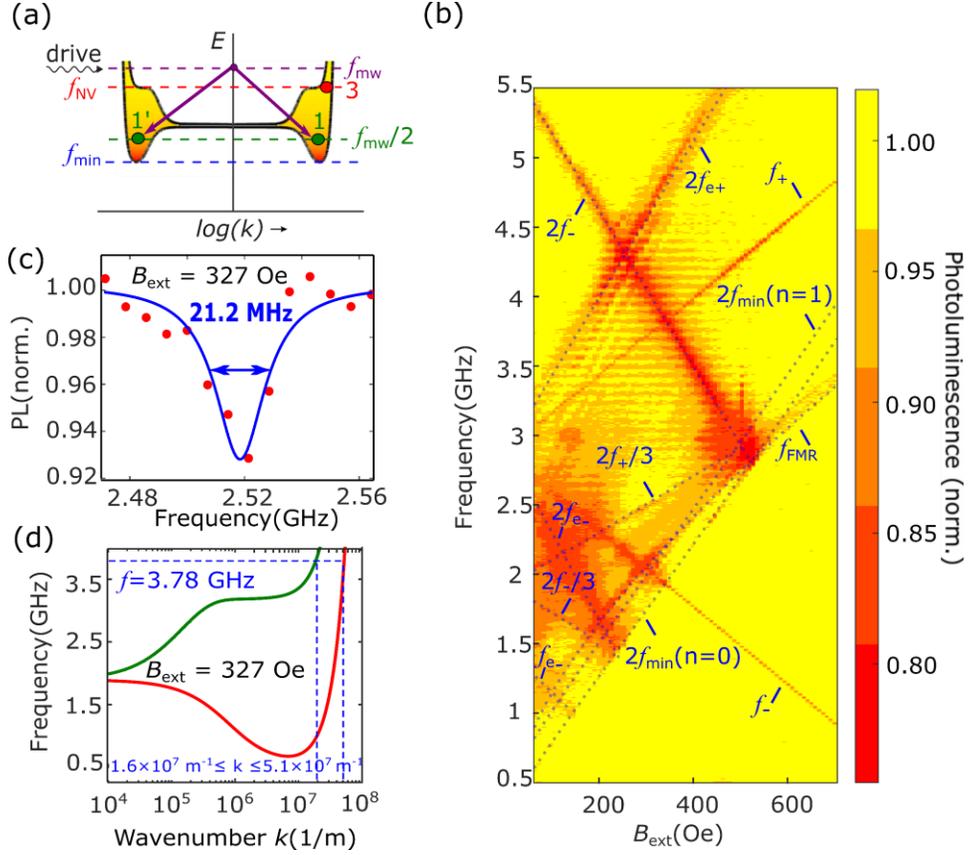

**Figure 3:** (a) Sketch of the magnon band structure of a 3-μm-thick YIG thin film, where the location of the microwave drive frequency $f_{mw}$ (purple color), the half frequency of the drive field $f_{mw}/2$ (green color), the NV ESR frequency $f_{NV}$ (red color), and the minimal magnon band energy $f_{min}$ (blue color) are marked. The upper branch of the dispersion curve corresponds to the magnetostatic surface spin wave mode and the lower branch corresponds to the backward volume spin wave mode. The shaded area between the two curves represents the magnon density. (b) Normalized PL intensity of the NV center as a function of $B_{ext}$ and the microwave frequency. Semi-transparent dashed lines represent the calculated dispersion curves of individual spin wave modes with the characteristic frequencies marked. (c) A linecut at $B_{ext}$ = 327 Oe showing the measured spin wave mode with a characteristic frequency: $f_{mw} = 2f_+/3$, from which the full width at half maximum $\Delta f_{mw}$ is determined to be 21.2 MHz. (d) The calculated magnon dispersion curves of the 3-μm-thick YIG thin film when $B_{ext}$ = 327 Oe. The red and green lines correspond to the situations where the magnon wavevector is parallel and perpendicular to the in-plane projection of the YIG magnetization, respectively. At $f_{mw}$ = 3.78 GHz, the estimated wavevector ranges from $1.6 \times 10^7$ m$^{-1}$ to $5.1 \times 10^7$ m$^{-1}$.

  Next, we comment on the accessible measurement range and resolution of magnon wavevectors that can be addressed by the NV single-spin sensors. Figure 3(c) shows a linecut at $B_{ext}$ = 327 Oe for the branch of spin wave mode with a characteristic frequency $f_{mw} = 2f_+/3$. The resonant condition happens at a microwave frequency $f_{mw}$ = 2.52 GHz with a full width at half maximum $\Delta f_{mw}$ = 21.2 MHz. The frequency of the generated magnons $f_m = 3f_{mw}/2$, equaling 3.78 ± 0.02 GHz. Based on the magnon dispersion curves calculated when the wavevector is parallel and perpendicular to the in-plane projection of the YIG magnetization,[43,45] the wavevector of such spin wave mode lies in the range of $1.6 \times 10^7$ m$^{-1}$ ≤ $k$ ≤ $5.1 \times 10^7$ m$^{-1}$ as shown in Fig. 3(d) (see Supporting Information for details). We note that this value is out of the measurement range of the conventional FMR spectroscopy and spin pumping techniques, and approaches the measurement limit of Brillouin light scattering (BLS). For FMR and spin pumping measurements, the accessible wavevectors of spin wave modes is mainly determined by the geometric dimensions of the on-chip microwave antenna, which typically lie in the range of $0 \leq k \leq 2.0 \times 10^7$ m$^{-1}$.[46] For BLS spectroscopy, the upper limit of the accessible wavevector is given by the Bragg relation: $k_{BLS} = n_R \frac{4\pi}{\lambda}$, where $n_R$ is the index of refraction of the YIG thin film. When $\lambda$ = 532 nm and $n_R$ = 2.34, $k_{BLS}$ is estimated to be ~$5 \times 10^7$ m$^{-1}$.[17,19,47] For the presented NV-based measurement platform, the detection sensitivity of magnons peaks at $k_{NV} = \frac{1}{d}$, where $d$ is the NV-to-sample distance.[28]. By employing the shallowly implanted NV centers or scanning NV microscopy,[48–50] $d$ could ultimately reach a regime of tens-of-nanometer, enabling a broadband detection of magnons with wavevectors up to ~$10^8$ m$^{-1}$. Note that the measured $\Delta f_{mw}$ shown in Fig. 3(c) is larger than the optimal resolution of NV spins, which is due to laser and microwave power induced broadening of the NV ESR linewidth.[51] The ultimate frequency resolution of an NV single-spin sensor is determined by its coherence time $T_2^*$.[25] Based on the Heisenberg uncertainty relationship $\Delta E = h/4\pi T_2^* = \Delta f h$ ($h$ is the Planck constant), when $T_2^*$ = 1.3 μs (see Supporting Information for details), the calculated frequency resolution $\Delta f$ can reach 0.06 MHz, yielding $\Delta k$ = 2000 m$^{-1}$ (when $k = 10^7$ m$^{-1}$).

Lastly, to illustrate the versatility of NV centers in accessing magnons with a broad range of wavevectors, we patterned the 100-nm-thick YIG thin film into a microdisk with a radius of 5 μm. A diamond nanobeam containing individual NV centers was transferred on top of the patterned microdisk as shown by the confocal image in Fig. 4(a). Due to the finite size effect of the magnetic material, only discrete values of the wavevectors $k = N\pi/R$ are allowed, where $N$ is an integer and $R$ is the radius of the patterned microdisk.[52,53] Figure 4(c) shows the calculated wavenumbers of two branches of discrete magnetostatic spin wave modes. The upper set of data corresponds to the magnetostatic surface spin wave with a propagation direction perpendicular to the in-plane projection of the YIG magnetization ($k \perp M$). The lower set corresponds to the backward volume spin wave with a propagation direction parallel to the in-plane projection of the magnetization ($k // M$).[43] In comparison to the parametric spin excitation discussed above, the wavevectors of the generated magnons are orders of magnitude smaller ranging from $6.3 \times 10^5$ to $1.9 \times 10^6$ m$^{-1}$. The measured ODMR map in Fig. 4(d) shows a set of spin wave features, whose field-dependent resonant frequencies are in agreement with the calculated magnetostatic surface spin wave modes with $N$ = 1, 2, and 3, respectively (see Supporting Information for details). The spatial profiles of the corresponding spin wave modes simulated via FastMag micromagnetic solver[54] are shown in Figs. 4(e) to 4(h). The obtained wavevectors are in quantitative agreement with the theoretical values.

In summary, we have demonstrated nanoscale detection of magnons via NV single-spin sensors. By tuning the thickness and the dimensions of the YIG thin films, the wavevectors of the excited magnons cover a broad range up to $5.1 \times 10^7$ m$^{-1}$. Detailed information, such as parametric spin pumping and the magnon band structure can be extracted via the performed ODMR measurements. The demonstrated coupling between the exchange magnons with high wavevectors and the NV single-spin sensors may also find applications in building NV-magnon-based hybrid quantum architectures for next-generation quantum information technologies, where long-range spin-entanglement and spin-wave-mediated control of the NV quantum spin states can be realized.

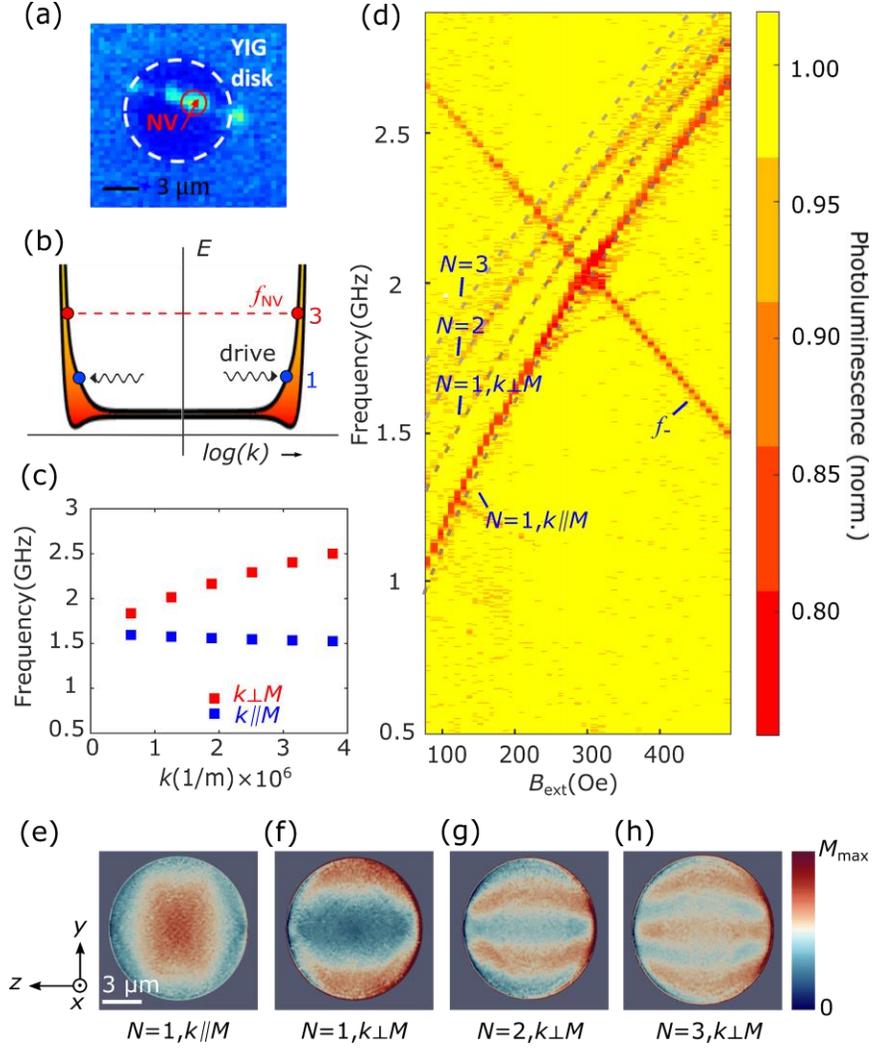

**Figure 4:** (a) Confocal image of a diamond nanobeam transferred onto a patterned 100-nm-thick YIG microdisk with a radius of 5 μm. The white dashed line marks the boundary of the YIG disk. (b) Sketch of the magnon band structure of the patterned YIG microstructure. The upper branch of the dispersion curve corresponds to the magnetostatic surface spin wave mode and the lower branch corresponds to the backward volume spin wave mode. The shaded area between the two curves represents the magnon density. (c) Discrete magnon frequencies as a function of the wavevectors calculated for the YIG microdisk. The red points correspond to the magnetostatic surface spin wave mode and the blue points correspond to the backward volume magnetostatic spin wave mode. (d) Normalized PL intensity of the NV center as a function of $B_{ext}$ and the microwave frequency. The calculated resonant frequencies as a function of the external magnetic field (semi-transparent dotted lines) agree with the observed features of the ODMR map. Micromagnetic simulations of the spatial profile of the transverse magnetization $M_T$ of the YIG disk with a microwave frequency $f_{mw}$ of (e) 1.64 GHz, (f) 1.83 GHz, (g) 2.02 GHz, and (h) 2.17 GHz, respectively. The external magnetic field $B_{ext}$ (200 Oe) is applied in $x$-$z$ plane with a 61 degree relative to the $x$-axis, and the microwave magnetic field is along the $x$-axis.

**Acknowledgement:**
The authors would like to thank Hanfeng Wang, Moyuan Chen, Albert Suceava, and Tony Zhou for help with experiments. We thank Vitaliy Lomakin and Lana Volvach for help with micromagnetic simulations. C. R. D. acknowledges support from a startup grant provided by UCSD. T. v. d. S. acknowledges support from the Dutch Research Council (NWO, Projectruimte grant 680.91.115). R. X, T. v. d. S and A. Y are supported by ARO Grant Number W911NF-17-1-0023.